\documentclass[11pt]{article}

\usepackage{amsmath}
\usepackage{graphicx}
\usepackage{latexsym}
\usepackage{amssymb}
\usepackage{amsbsy}

\DeclareSymbolFont{msbm}{U}{msb}{m}{n}
\DeclareMathSymbol{\C}{\mathalpha}{msbm}{'103}
\DeclareMathSymbol{\R}{\mathalpha}{msbm}{'122}
\DeclareMathSymbol{\Z}{\mathalpha}{msbm}{'132}
\DeclareMathSymbol{\N}{\mathalpha}{msbm}{'116}

\newtheorem{remark}{Remark}

\def\RR{\mathbb R}

\def\be{\begin{equation}}
\def\ee{\end{equation}}
\def\bea{\begin{eqnarray}}
\def\ba{\begin{array}{l}\displaystyle}
\def\eea{\end{eqnarray}}
\def\ea{\end{array}}

\begin{document}
\title{Mean field mutation dynamics and the continuous Luria-Delbr\"uck distribution}

\author{Eugene Kashdan\thanks{Applied Mathematics Department, Tel Aviv University, Israel ({\tt
ekashdan@post.tau.ac.il}).}\and Lorenzo
Pareschi\thanks{Mathematics Department, University of Ferrara,
Italy ({\tt lorenzo.pareschi@unife.it}).} }
\maketitle

\begin{abstract}
The Luria-Delbr\"uck mutation model has a long history and has
been mathematically formulated in several different ways. Here we
tackle the problem in the case of a continuous distribution using
some mathematical tools from nonlinear statistical physics.
Starting from the classical formulations we derive the
corresponding differential models and show that under a suitable
mean field scaling they correspond to generalized Fokker-Planck equations for
the mutants distribution whose solutions are given by the
corresponding Luria-Delbr\"uck distribution.
Numerical
results confirming the theoretical analysis are also presented.
\end{abstract}

\maketitle

{\bf Keywords:} Luria-Delbr\"uck distribution, kinetic models,
mutation rates, Fokker-Planck equations


\section{Introduction}

Application of tools from nonlinear statistical physics to
describe different biological phenomena started to gain popularity
in the recent years and it represents one of the major challenges
in contemporary mathematical modeling \cite{Bell,BD, HBSSD, MD}. In
particular, powerful methods borrowed from kinetic theory of
rarefied gases have been successfully employed to construct master
equations of Boltzmann type, usually referred to as kinetic
equations, describing the emergency of universal behaviors through
their equilibria \cite{CPT,NPT,PT1}.

An important example of emergent behavior describes building of
tumors by cancer cells and their migration through the tissues
\cite{BD, Grizzi, HBSSD, MD}. Another famous example to consider in this contest
is the classical Luria-Delbr\"uck mutation problem \cite{Luria, Zheng}.
Certainly, the basic entities in these examples differ from the
physical particles in that they already have an intermediate
complexity themselves. However, for some specific phenomena, the
statistical behavior of the system is related to the peculiar way
entities interact and not to their internal complex structure.

First experimental and theoretical analysis of mutation process in
bacteria was published by Luria and  Delbr\"uck in 1943
\cite{Luria} (Nobel Prize in Physiology and Medicine, 1969). The
goal of the study conducted by Luria and Delbr\"uck was to
estimate the mutation rate in bacterial populations by observing
the fraction of the final cells that carry a mutation. In their
experiment, often called {\em fluctuation test}, they have shown
that in bacteria, genetic mutations arise in the absence of
selection, rather than being a response to selection. They
explained these results mathematically by the occurrence of a
constant rate of random mutations in each generation of bacteria
growing in the initial culture.

The distribution of the number of mutants in a culture that has
been grown under conditions in which these mutations did not
confer a selective advantage to the cells bearing them, came to be
known as the {\it Luria-Delbr\"uck distribution}.

In multicellular organisms, connection between mutagenesis and
carcinogenesis is broadly accepted (see, for instance, \cite{NBG})
and the Luria-Delbr\"uck  distribution plays an important role in
the study of cancer, because tumor progression depends on how
heritable changes (mutations) accumulate in cell lineages. In
particular, the Luria-Delbr\"uck model focuses on the distribution
of mutations in an exponentially expanding clone of cells. We
should mention here that several models describing the
Luria-Delbr\"uck experiment have been proposed. The most famous is
the Lea-Coulson formulation \cite{Lea} based on the introduction
of a random rate of mutations instead of constant rate implemented
in \cite{Luria}. Other formulations have been introduced in
\cite{Bar,Arm,Zheng3}. We refer the reader to \cite{Zheng} for a
comprehensive review of the subject. The mathematics behind the
Luria-Delbr\"uck experiment (most often in the Lea-Coulson
formulation) has been studied by several authors
\cite{Ang,JTR,Kep1,Kep2,Mandelbrot,Zheng2}.

The specific shape of Luria-Delbr\"uck distribution is strictly
connected to the particular formulation of the mutation dynamics.
In general, no analytic expressions of such distributions are
available, however, exact expressions may be available for the
characteristic functions, the probability generating functions and
the moments. This indeed is the case of both the original
(Luria-Delbr\"uck) and the Lea-Coulson formulations considered in
this manuscript. For such a reason a lot of effort has been
invested in the development of computational methods and
asymptotic approximations of Luria-Delbr\"uck distribution (see
\cite{Zheng} and the references therein).

Here we introduce kinetic master equations, in the spirit of
\cite{PT1}, where the number of mutants is considered as a
continuous variable instead of a discrete one. The behavior of the
resulting differential models are then studied by using a suitable
mean field asymptotic scaling, which permits to recover the exact moments of
the original formulations. As the limit we obtain generalized
Fokker-Planck equations and, for the classical Luria-Delbr\"uck setting, show
that the long time behavior is characterized by the corresponding Luria-Delbr\"uck distribution.
The same evidence is give numerically also for the Lea-Coulson setting.

The rest of the paper is organized as follows. In the next section
we consider the standard Luria-Delbr\"uck experimental settings and
present the corresponding kinetic master equations. Then we introduce a mean field scaling that
permits to derive a generalized Fokker-Planck model whose solution is given by
the original Luria-Delbr\"uck distribution of mutants. In section
3 we consider the case of a variable mutation rate as proposed by
the Lea-Coulson formulation. Derivation of the
generalized Fokker-Planck model for this case is also presented and its solution
is discussed. Simple diffusion approximations of the mutation process
are also derived in this section. Then, in section 4, we present numerical
experiments confirming our theoretical analysis. Concluding
remarks and possible extensions of this work are reported in the last section.

\section{The Luria-Delbr\"uck formulation}
The mathematical formulation of the Luria-Delbr\"uck model is based on the following assumptions
\begin{itemize}
\item The process starts at time $t=0$ with a number $N_0$ of normal cells and no mutants.
\item Normal cells replicate at a constant rate $\alpha$.
\item Mutation occur randomly at a rate characterized by a Poisson process with intensity proportional to
the total number of normal cells. We denote with $\mu$ the
mutation rate per cell.
\item The offspring of every mutant cell replicate at a constant rate $\beta$.
\end{itemize}
As a consequence of the above assumptions if we denote by $N(t)$
the number of normal cells we obtain the differential problem \be
\frac{d N(t)}{dt}=\alpha N(t)-\mu N(t),\quad N(0)=N_0,
\label{eq:nor}\ee which gives \be N(t)=N_0 e^{(\alpha-\mu)t},\quad
t > 0. \ee The fundamental question in the above model is the
estimation of the distribution of the number of mutants in time.
Such distribution, in fact, is of vital importance when estimating
the mutation rates. We shall denote by $f(m,t)$ the probability
density of having $m$ mutant cells at time $t$.

\subsection{A kinetic model for mutations} In the sequel we will
assume that the random variable $m$ characterizing the number of
mutants at time $t$ takes values in $\RR^{+}$. Such an assumption,
as we will see later, permits us to tackle the problem with some tools
borrowed from kinetic theory \cite{Ce, CPT, NPT, PT1}. Thus we
have
\[
\int_{\RR^+} f(m,t)\,dm = 1,
\]
and denote by $M(t)$ the expected number of mutant cells at time
$t$ \be M(t)=\int_{\RR^+} f(m,t)m\,dm. \ee We consider the
following microscopic dynamic for the random variable $m$ \be
m'=m+\beta m+\eta, \ee where $\eta\geq 0$ (absence of backward
mutations) is a discrete random variable distributed accordingly
to a Poisson density $p(N(t),\eta)$ with mean $\mu N(t)$.

By standard arguments we can write the following
kinetic master equation for the time evolution of the distribution
of mutant cells \be \frac{\partial f(m,t)}{\partial t} =
\sum_{i=0}^{\infty} \left[B_{'m\to m}(N(t),\eta_i)
\frac1{J}f('m,t)-B_{m\to m'}(N(t),\eta_i) f(m,t)\right], \ee where
$'m=(m-\eta_i)/J$, $J=1+\beta$, and the transition rates are given
by
\[
B_{'m\to m}(N(t),\eta_i) = p(N(t),\eta_i)\chi('m \geq 0),\quad
B_{m\to m'}(N(t),\eta_i) = p(N(t),\eta_i)\chi(m' \geq 0),
\]
with $\chi(I)$ the indicator function of the set $I$. The above
equation is complemented with the initial data \be f(m,0)=f_0(m).
\ee

Since the random variable $\eta_i$ takes values only in $\RR^+$
the weak form of the kinetic equation yields the simpler
representation \be \frac{\partial}{\partial
t}\int_{\RR^+}f(m,t)\varphi(m)\,dm = \sum_{i=0}^{\infty}
p(N(t),\eta_i)\int_{\RR^+} f(m,t)[\varphi(m')-\varphi(m)]\,dm,
\label{eq:km} \ee where $\varphi(m)$ is a smooth function.

It is easy to see that taking $\varphi(m)=1$ we get the
consistency condition \be \frac{\partial}{\partial
t}\int_{\RR^+}f(m,t)\,dm=0. \ee The choice $\varphi(m)=m$ yields
the evolution of the mean number of mutant cells and gives \[
\frac{\partial}{\partial t}\int_{\RR^+}f(m,t)m\,dm =
\sum_{i=0}^{\infty}p(N(t),\eta_i) \int_{\RR^+} f(m,t)[\beta m +
\eta_i]\,dm=\beta\int_{\RR^+}f(m,t)m\,dm-\mu N(t), \] which
corresponds to the simple ODE \be \frac{dM(t)}{dt}=\beta M(t)+\mu
N_0e^{(\alpha-\mu)t}. \label{eq:mut}\ee The above equation can be
solved exactly and, starting with initial data $M(0)=M_0$, gives
\be M(t)=\left\{
       \begin{array}{ll}
         \displaystyle\frac{\mu}{\alpha-\mu-\beta}N_0 e^{(\alpha-\mu) t}(1-e^{-(\alpha-\mu-\beta)t})+M_0e^{\beta t}, & \beta \neq \alpha-\mu,\\[+.5cm]
         \mu N_0 t e^{(\alpha-\mu) t}, & \beta=\alpha-\mu.
       \end{array}
     \right.
     \label{solm}
\ee

Note that the above expression for the mean coincides with what
found in the literature for the Luria-Delbr\"uck distribution
\cite{Zheng}.

Similarly we can compute the evolution of the variance. Starting
from $\varphi(m)=m^2$ we have for the second moment $M_2(t)$ the
equation
\begin{eqnarray*}
\frac{\partial}{\partial t}\int_{\RR^+}f(m,t)m^2\,dm &=& \sum_{i=0}^{\infty} p(N(t),\eta_i)\int_{\RR^+} f(m,t)[\beta(2+\beta) m^2 + \eta_i^2+2(1+\beta)m\eta_i]\,dm,\\
&=& \beta(2+\beta)\int_{\RR^+}f(m,t)m^2\,dm \\
&+&<\eta^2>+2(1+\beta) \mu N(t)\int_{\RR^+}f(m,t)m\,dm,
\end{eqnarray*}
where $<\eta^2>=\mu N(t)(1+\mu N(t))$ denotes the second moment of $\eta_i$.\\
Thus the variance
\[
V(t)=\int_{\RR^+}f(m,t)m^2\,dm-M(t)^2,
\]
follows the differential equation \be \frac{dV(t)}{dt}=2\beta
V(t)+\beta^2 M_2(t)+\mu N(t)(1+2\beta M(t)+\mu N(t)),
\label{eq:varm}\ee which differs from the variance of the
Luria-Delbr\"uck distribution. The exact analytical solution to
(\ref{eq:varm}) can be easily computed, we omit it for brevity.

To analyze the long time behavior let us consider the case
$N_0=1$, $M_0=0$, which represents the situation, where the mean
numbers of normal cells and mutant cells are initially one and
zero respectively (a standard Luria-Delbr\"{u}ck setup
\cite{Luria}). Note that, when the growth rates are identical
$\alpha=\beta$, the average total number of cells is
$B(t)=M(t)+N(t)=e^{\beta t}$ which is simply a birth process at
rate $\beta$.

Let us define next the concentration of the mutant cells as
$\rho(t)={M(t)}/{B(t)}$. The long time behavior of $\rho$
according to \eqref{solm} is given by
\begin{align}
\label{ltb} \lim_{t\rightarrow\infty} \rho(t)=
\begin{cases}
1, ~~& \beta\geq\alpha-\mu\;,\\
\displaystyle\frac{\mu}{\alpha-\beta}, ~~ &\beta<\alpha-\mu
\end{cases}
\end{align}

\begin{figure}[t]
\begin{center}
\includegraphics[width=11cm,height=7cm]{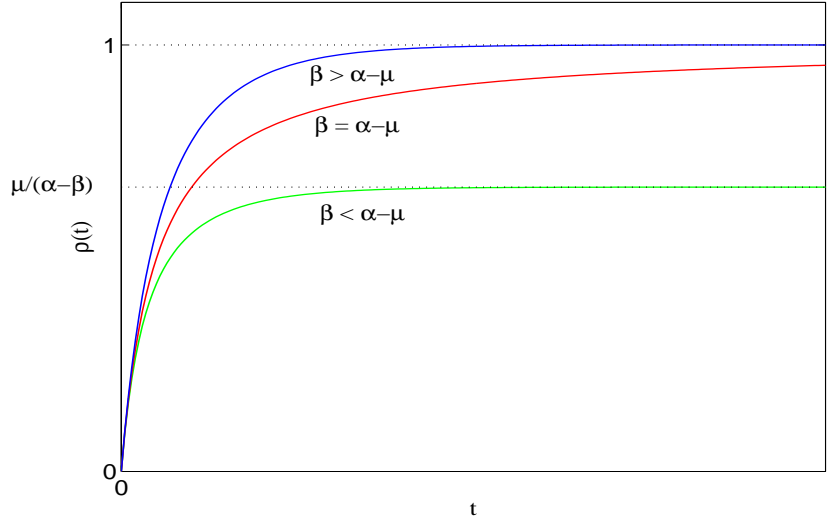}
\caption{Long time behavior of mutant cell concentration as
function of time}
\end{center}
\label{fg:mp}
\end{figure}

The behavior is sketched in Figure 1. Mutants become predominant
when $\beta\geq\alpha-\mu$. It is interesting to revise that the
asymptotic behavior of $\rho$ shows exponential convergence for
all values of $\beta$ except for $\beta=\alpha-\mu$, where we have
\begin{align*}
\rho(t)=\frac{\mu t}{1+\mu t}
\end{align*}

For $\beta<\alpha-\mu$ mutant cells represent a fraction
$\mu/(\alpha-\beta)$ of the population and no predominant behavior
is observed.

\subsection{Derivation of the generalized Fokker-Planck model}
\label{ss2}
In order to study the asymptotic behavior of the model
we introduce the following scaling. We set \be{\tilde
f}(m,\tau)=f(m,t),\qquad \tau = \varepsilon t. \label{eq:sc1}\ee
We are interested in the behavior for small values of
$\varepsilon$ such that \be \lim_{\varepsilon,\beta\to 0}
\frac{\beta}{\varepsilon}=\gamma,\quad
\lim_{\varepsilon,\beta_1\to 0}
\frac{\beta_1}{\varepsilon}=\gamma_1\quad \lim_{\varepsilon,\mu\to
0}\frac{\mu}{\varepsilon}=\nu, \label{eq:sc}\ee where
$\beta_1=\alpha-\mu$.

Roughly speaking the above scaling limit implies $m'\approx m$ and
corresponds to the realistic assumption of considering the
the birth-mutation dynamic as a slow process originated by many small
variations of the number of mutants. For this reason we will refer
to the limiting dynamic as a mean field dynamic. In particular, as we will
show, it is consistent with the evolution of the variance of the
original Luria-Delbr\"uck model.

To this aim, let us first point out that such scaling does not
affect the evolution of the time rescaled mean $\tilde M(\tau)$
which in the limit (\ref{eq:sc}) is governed by \be
\frac{d\tilde{M}(\tau)}{d\tau}=\gamma \tilde{M}(\tau)+\nu
N_0e^{\gamma_1\tau}. \label{eq:tsm}\ee For the time rescaled
variance $\tilde V(\tau)$ in the limit $\varepsilon\to 0$ we
obtain the simpler equation \be
\frac{d\tilde{V}(\tau)}{d\tau}=2\gamma \tilde{V}(\tau)+\nu
N_0e^{\gamma_1\tau}. \ee Clearly the above equation can be solved
exactly, and for $\tilde V(0)=0$ gives \be \tilde V(\tau)=\left\{
       \begin{array}{ll}
         \displaystyle\frac{\nu}{\gamma_1-2\gamma}N_0 e^{2\gamma\tau}(e^{(\gamma_1-2\gamma)\tau}-1), & 2\gamma \neq \gamma_1,\\[+.5cm]
         \mu N_0 \tau e^{\gamma_1\tau}, & 2\gamma=\gamma_1,
       \end{array}
     \right.
\ee which is the same expression of the variance for the
Luria-Delbr\"uck distribution \cite{Zheng}.

The time scaled distribution $\tilde f(m,\tau)$ satisfies the
equation
\begin{eqnarray}
\frac{\partial}{\partial \tau}\int_{\RR^+}\tilde
f(m,\tau)\varphi(m)\,dm=
\frac1{\varepsilon}\sum_{i=0}^{\infty}p({\tilde
N}(\tau),\eta_i)\int_{\RR^+} \tilde
f(m,\tau)[\varphi(m')-\varphi(m)]\,dm. \label{eq:km2}
\end{eqnarray}
Let us expand $\varphi(m')$ in Taylor series
\[
\varphi(m')=\varphi(m)+\sum_{k=1}^{\infty}
\frac{(m'-m)^k}{k!}\varphi^{(k)}(m),
\]
and substituting into (\ref{eq:km2}) we have
\begin{eqnarray}
\frac{\partial}{\partial \tau}\int_{\RR^+}\tilde
f(m,\tau)\varphi(m)\,dm =
\frac1{\varepsilon}\sum_{k=1}^{\infty}\frac1{k!}
\sum_{i=0}^{\infty}p({\tilde N}(\tau),\eta_i)\int_{\RR^+} \tilde
f(m,\tau)(\beta m+\eta_i)^k\varphi^{(k)}(m)\,dm.
\end{eqnarray}
Using the binomial formula we can write
\[
(\beta m +\eta_i)^k=\sum_{j=0}^k \left({k\atop j}\right)\beta^j
m^j \eta_i^{k-j},
\]
and then \begin{eqnarray*} \frac{\partial}{\partial
\tau}\int_{\RR^+}\tilde f(m,\tau)\varphi(m)\,dm
=\sum_{k=1}^{\infty}\frac1{k!}\sum_{j=0}^k \left({k\atop j}\right)
\frac{\beta^j <\eta^{k-j}>}{\varepsilon} \int_{\RR^+}\tilde
f(m,\tau) m^j \varphi^{(k)}(m)\,dm,  \end{eqnarray*} where
$<\eta^{k-j}>$ denotes the $(k-j)$-th moment of $\eta$.

We are interested in taking the limit $\varepsilon,
\beta_1,\beta,\mu\to 0$. Now since $\eta_i$ is given by a Poisson
process
\[
<\eta^{k-j}>=\sum_{i=0}^{\infty}
p(\tilde{N}(\tau),\eta_i)\eta_i^{k-j}=\mu
\tilde{N}(\tau)+O(\mu^2),
\]
and thus
\[
\lim_{\varepsilon,\mu\to 0} \frac{<\eta^{k-j}>}{\varepsilon} =
\nu\tilde{N}(\tau),\quad \forall\, j < k.
\]
In the limit we finally obtain \begin{eqnarray*}
\frac{\partial}{\partial \tau}\int_{\RR^+}\tilde
f(m,\tau)\varphi(m)\,dm -\gamma\int_{\RR^+} m \tilde
f(m,\tau)\varphi'(m)\,dm=\sum_{k=1}^{\infty}\frac{\nu}{k!}{\tilde
N(\tau)} \int_{\RR^+}\tilde f(m,\tau) \varphi^{(k)}(m)\,dm,
\end{eqnarray*}
which corresponds to the generalized Fokker-Planck equation \be
\frac{\partial}{\partial \tau} \tilde
f(m,\tau)+\gamma\frac{\partial}{\partial m}(m \tilde
f(m,\tau))=\nu {\tilde N(\tau)} {\cal L}_{KM}(\tilde f (m,\tau)),
\label{eq:gfp} \ee where \be {\cal L}_{KM}(\tilde f(m,\tau)):=
\sum_{k=1}^{\infty}\frac{(-1)^k}{k!}\frac{\partial^{(k)}}{\partial
m^{(k)}}\tilde f(m,\tau)\label{eq:kmo}\ee is the Kramers-Moyal
operator.\\
Let us scale our solution accordingly with \be
g(m,\tau)=u(\tau){\tilde f}(mu(\tau),\tau),\quad
u(\tau)=e^{\gamma\tau}. \ee Then $g(m,\tau)$ satisfies the
equation \be \frac{\partial}{\partial \tau} g(m,\tau)=\nu {\tilde
N(\tau)} \sum_{k=1}^{\infty}\frac{(-1)^k
e^{-k\gamma\tau}}{k!}\frac{\partial^{(k)}}{\partial
m^{(k)}}g(m,\tau). \label{eq:gfps2} \ee Now, we introduce the
Fourier transform
\[
{\hat g}(\xi,\tau)=\int_{\RR} g(m,\tau)e^{-im\xi}\,dm,
\]
to obtain \begin{eqnarray*} \frac{\partial}{\partial \tau} {\hat
g}(\xi,\tau)&=&\nu {\tilde N(\tau)}{\hat g}(\xi,\tau)
\sum_{k=1}^{\infty}\frac{(-1)^k}{k!}(i\xi e^{-\gamma\tau})^k\\
&=& \nu N_0 e^{\tau}{\hat g}(\xi,\tau) (e^{-i\xi
e^{-\gamma\tau}}-1).
\end{eqnarray*}
The solution of the above differential equation can be written in
the form \be {\hat g}(\xi,\tau)={\hat g}_0(\xi)\exp\left(\nu N_0
\int_{0}^{\tau} \left(e^{-i\xi e^{-\gamma
z}}-1\right)e^z\,dz\right), \label{eq:LDG} \ee with
\[ {\hat g}_0(\xi)=\int_{\RR} g_0(m)e^{-im\xi}\,dm. \]
Note that equation (\ref{eq:LDG}) is exactly the characteristic
function of the Luria-Delbr\"uck distribution \cite{Zheng}. This
shows that the solution of the kinetic model (\ref{eq:km}), in the
asymptotic limit described by equations
(\ref{eq:sc1})-(\ref{eq:sc}), coincides with the classical
Luria-Delbr\"uck distribution characterized by (\ref{eq:LDG}).

\section{The Lea-Coulson setting}
The original formulation by Luria and Delbr\"uck assumed a
deterministic growth rate for mutant cells, which seemed too
stringent an assumption to allow for efficient extraction of
reliable information about mutation rates from experimental data
\cite{Zheng}. A slightly different mathematical formulation was
proposed by Lea and Coulson \cite{Lea}, who adopted a preferential
attachment process (or Yule process) with birth rate $\beta$ to
describe the growth of mutant cells. Today the Lea-Coulson
formulation is the prominent model for the study of mutation rates
and is now commonly referred to as the Luria-Delbr\"uck model.

\subsection{The kinetic equation}
The formulation of the Lea-Coulson assumptions is based on
considering the microscopic dynamic \be m'=m+\vartheta+\eta, \ee
where now $\vartheta\geq 0$ is a discrete random variable
distributed accordingly to a Poisson density $p(m,\vartheta)$ with
intensity $\beta m$.

The weak-form of the kinetic equation now reads \be
\frac{\partial}{\partial t}\int_{\RR^+}f(m,t)\varphi(m)\,dm =
\sum_{i,j=0}^{\infty}
p(N(t),\eta_i)\int_{\RR^+}p(m,\vartheta_j)f(m,t)[\varphi(m')-\varphi(m)]\,dm,
\label{eq:km2c} \ee with $\varphi(m)$ a smooth function and we
used a compact notation to denote the double sum over $i$ and $j$.

Again taking $\varphi(m)=1$ we have that $f(m,t)$ is a p.d.f. for
all times, whereas the choice $\varphi(m)=m$ yields the evolution
of the mean number of mutant cells. It is a simple computation to
show that we get the same equation for the mean as in classical
Luria-Delbr\"uck setting since
\begin{eqnarray*}
 \frac{\partial}{\partial t}\int_{\RR^+}f(m,t)m\,dm &=&
\sum_{i,j=0}^{\infty} p(N(t),\eta_i)\int_{\RR^+}
p(m,\vartheta_j)[\vartheta_j +
\eta_i]f(m,t)\,dm\\
&=&\beta M(t)+\mu N(t).
\end{eqnarray*}
Let us now compute the evolution of the variance. Taking
$\varphi(m)=m^2$ we have for the second moment $M_2(t)$

\begin{eqnarray*}
\frac{\partial}{\partial t}\int_{\RR^+}f(m,t)m^2\,dm &=&
\sum_{i,j=0}^{\infty}
p(N(t),\eta_i) \int_{\RR^+}p(m,\vartheta_j)  f(m,t)[\vartheta_j^2 + \eta_i^2+2\eta_i\vartheta_j \\&+&2m(\eta_i+\vartheta_j)]\,dm,\\
&=& <\eta^2>+M(t)(\beta+2\beta\mu N(t)+2\mu
N(t))\\&+&\beta(2+\beta)\int_{\RR^+}f(m,t)m^2\,dm,
\end{eqnarray*}
which gives the following expression for the kinetic variance \be
\frac{dV(t)}{dt}=2\beta V(t)+\beta^2 M_2(t)+\beta M(t)(1-2\mu
N(t))+\mu N(t)(1+\mu N(t)). \label{eq:varm2b} \ee The above
expression coincides with (\ref{eq:varm}) except for the presence
of the additional term $\beta M(t)$ on the r.h.s. As before we
omit for brevity the exact solution to (\ref{eq:varm2b}).

\subsection{The generalized Fokker-Planck limit}
Following the analysis of section \ref{ss2} we can introduce the
mean field scaling defined by (\ref{eq:sc1}) and (\ref{eq:sc}).

Again the scaling does not affect the evolution of the time
rescaled mean $\tilde M(\tau)$ which in the asymptotic limit is
governed again by (\ref{eq:tsm}).

Now the time rescaled variance $\tilde V(\tau)$ when
$\varepsilon\to 0$ is governed by \be
\frac{d\tilde{V}(\tau)}{d\tau}=2\gamma
\tilde{V}(\tau)+\gamma\tilde M(\tau)+\nu N_0e^{\gamma_1\tau}. \ee
The exact solution of the above equation for $\tilde V(0)=0$
yields the classical expression of the variance in the Lea-Coulson
formulation \cite{Zheng} \be \tilde V(\tau)=\left\{
       \begin{array}{ll}
         \displaystyle\frac{\nu}{\gamma_1-\gamma}N_0 e^{2\gamma\tau}\left[\frac{\gamma_1(e^{(\gamma_1-2\gamma)\tau}-1)}{\gamma_1-2\gamma}+(e^{-\gamma\tau}-1)\right], & 2\gamma \neq \gamma_1,\,\gamma\neq\gamma_1\\[+.5cm]
         \displaystyle\frac{\nu}{\gamma}N_0 e^{\gamma\tau}(1-e^{\gamma\tau})+2\nu N_0 \tau e^{2\gamma\tau}, & 2\gamma = \gamma_1,\\[+.5cm]
         \displaystyle\frac{2\nu}{\gamma}N_0 e^{\gamma\tau}(e^{\gamma\tau}-1)-\nu N_0\tau e^{\gamma\tau}, & \gamma =
         \gamma_1.
       \end{array}
     \right.
     \label{eq:vlc}
\ee

Now let us consider the evolution of the rescaled probability
density function $\tilde f(m,\tau)$. We have
\begin{eqnarray}
\frac{\partial}{\partial \tau}\int_{\RR^+}\tilde
f(m,\tau)\varphi(m)\,dm= \frac1{\varepsilon} \sum_{i,j=0}^{\infty}
p(\tilde N(\tau),\eta_i)\int_{\RR^+}p(m,\vartheta_j) \tilde
f(m,\tau)[\varphi(m')-\varphi(m)]\,dm. \label{eq:km2d}
\end{eqnarray}
By the same method introduced in section \ref{ss2} we expand
$\varphi(m')$ in Taylor series and substitute into (\ref{eq:km2c})
to get
\begin{eqnarray*}
\frac{\partial}{\partial \tau}\int_{\RR^+}\tilde
f(m,\tau)\varphi(m)\,dm =
\frac1{\varepsilon}\sum_{k=1}^{\infty}\frac1{k!}
\sum_{i,j=0}^{\infty} p(\tilde
N(\tau),\eta_i)\int_{\RR^+}p(m,\vartheta_j)\tilde f(m,\tau)
(\vartheta_j+\eta_i)^k\varphi^{(k)}(m)\,dm.
\end{eqnarray*}
Using the binomial formula on $(\vartheta_j+\eta_i)^k$ we can
write
\begin{eqnarray*}
\frac{\partial}{\partial \tau}\int_{\RR^+}\tilde
f(m,\tau)\varphi(m)\,dm =\sum_{k=1}^{\infty}\frac1{k!}\sum_{h=0}^k
\left({k\atop h}\right) \frac{<\eta^{k-h}>}{\varepsilon}
\int_{\RR^+}<\vartheta_j>^h\tilde f(m,\tau) \varphi^{(k)}(m)\,dm,
\end{eqnarray*}
where $<\vartheta^h>$ and $<\eta^{k-h}>$ denote the $h$-th moment
and the $(k-h)$-th moment of $\vartheta$ and $\eta$ respectively.

We are interested in taking the limit $\varepsilon,
\beta_1,\beta,\mu\to 0$. Now since $\eta_i$ and $\vartheta_j$ are
given by Poisson processes
\[
<\eta^{k-h}>=\mu \tilde{N}(\tau)+O(\mu^2),\quad
<\vartheta^h>=\beta m +O(\beta^2),
\]
and then
\[
\lim_{\varepsilon,\mu,\beta\to 0}
\frac{<\eta^{k-h}><\vartheta^{h}>}{\varepsilon} = \left\{%
\begin{array}{ll}
    \nu\tilde{N}(\tau), & h=0; \\
    \gamma m, & h=k; \\
    0, & 0<h<k.
\end{array}\right.
\]
In the limit we finally obtain \begin{eqnarray*}
&&\frac{\partial}{\partial \tau}\int_{\RR^+}\tilde
f(m,\tau)\varphi(m)\,dm - \gamma\sum_{k=1}^{\infty}\frac{1}{k!}
\int_{\RR^+} m\tilde f(m,\tau) \varphi^{(k)}(m)\,dm\\&&=\nu\tilde
N(\tau)\sum_{k=1}^{\infty}\frac{1}{k!} \int_{\RR^+}\tilde
f(m,\tau) \varphi^{(k)}(m)\,dm,
\end{eqnarray*}
which corresponds to the generalized Fokker-Planck equation \be
\frac{\partial}{\partial \tau} \tilde f(m,\tau)={\cal
L}_{KM}((\gamma m + \nu {\tilde N(\tau)})\tilde f (m,\tau)),
\label{eq:gfp2b} \ee where ${\cal L}_{KM}(\cdot)$ is the
Kramers-Moyal operator given by (\ref{eq:kmo}).

\begin{remark}[A simplified model] A further explicit analysis for equation
(\ref{eq:gfp2b}) seems quite difficult. Let us remark that taking
the simplified microscopic dynamic \be m'=m+\vartheta+\eta, \ee
where now the growth of the random variable $m$ depends only on the
mean value of the variable itself, in weak-form we obtain the
kinetic equation \be \frac{\partial}{\partial
t}\int_{\RR^+}f(m,t)\varphi(m)\,dm = \sum_{i,j=0}^{\infty}
p(N(t),\eta_i)p(M(t),\vartheta_j)\int_{\RR^+}f(m,t)[\varphi(m')-\varphi(m)]\,dm.
\label{eq:km2a} \ee In the same scaling limit we obtain
\begin{eqnarray*} \frac{\partial}{\partial \tau}\int_{\RR^+}\tilde
f(m,\tau)\varphi(m)\,dm
=\sum_{k=1}^{\infty}\frac{1}{k!}\left({\nu\tilde
N(\tau)+\gamma\tilde M(\tau)}\right) \int_{\RR^+}\tilde f(m,\tau)
\varphi^{(k)}(m)\,dm,
\end{eqnarray*}
which corresponds to the generalized Fokker-Planck equation \be
\frac{\partial}{\partial \tau} \tilde f(m,\tau)=\left(\nu {\tilde
N(\tau)}+\gamma\tilde M(\tau)\right) {\cal L}_{KM}(\tilde f
(m,\tau)). \label{eq:gfp2} \ee
Setting $g(m,\tau)=\tilde
f(m,\tau)$ and passing to the Fourier transform we obtain readily
\begin{eqnarray*}
\frac{\partial}{\partial \tau} {\hat g}(\xi,\tau)&=&\left(\nu
{\tilde N(\tau)}+\gamma M(\tau)\right){\hat g}(\xi,\tau)
(e^{-i\xi}-1).
\end{eqnarray*}
The solution can be written in the form \be {\hat
g}(\xi,\tau)={\hat g}_0(\xi)\exp\left((e^{-i\xi}-1)
\int_{0}^{\tau} \left(\nu {\tilde N(z)}+\gamma \tilde
M(z)\right)\,dz\right), \label{eq:LCS} \ee which now can be
integrated exactly and gives the characteristic function \be {\hat
g}(\xi,\tau)=
    {\hat g}_0(\xi)\exp\left(\displaystyle
    (e^{-i\xi}-1)\tilde M(\tau)\right).
    \label{eq:LCG}
\ee Expression (\ref{eq:LCG}) coincides with the characteristic
function of a Poisson process with mean $\tilde M(\tau)$. Note
however that in our case $m$ is a continuous variable and so we
cannot conclude that the unknown mutant distribution has a Poisson
distribution but it would be rather characterized by some
continuous approximation of a Poisson distribution, like
incomplete Gamma functions \cite{Mar}.
\end{remark}

\subsection{Diffusion approximations} In order to compute the
distribution of mutants we can clearly use the characteristic
function (\ref{eq:LDG}) or (\ref{eq:LCG}). However if we are
interested in developing more realistic models where mutations are
time dependent and related to space variables, like in a multicellular
organism, we cannot rely on explicit solutions. Thus we must either
solve the kinetic models (\ref{eq:km}) and (\ref{eq:km2c}) (or
(\ref{eq:km2a})), under the scaling (\ref{eq:sc}), or the
corresponding generalized Fokker-Planck equations (\ref{eq:gfp})
and (\ref{eq:gfp2b}) (or (\ref{eq:gfp2})). In the latter case the
Kramers-Moyal operator (\ref{eq:gfp}) must be truncated at a
finite number of terms and solved numerically.

The Pawula theorem \cite{fep} states that truncations of the
generalized Fokker-Planck equation in the form (\ref{eq:gfp}) with
finite derivatives greater than two leads to a contradiction to
the positivity of the distribution function. It is then natural to
consider a second order truncation to (\ref{eq:gfp}) which in the
original variables corresponds to the following diffusion
approximation of the Luria-Delbr\"uck dynamic \be
\frac{\partial}{\partial \tau} \tilde
f(m,\tau)+\frac{\partial}{\partial m}[(\gamma m+\nu \tilde
N(\tau)) \tilde f(m,\tau)]=\frac12\nu \tilde
N(\tau)\frac{\partial^2}{\partial m^2} \tilde f(m,\tau).
\label{eq:fp1}\ee It is immediate to verify that the solution to
the above equation has the same mean and variance as the original
Luria-Delbr\"uck distribution.

Similarly we obtain the following diffusion approximation of the
Lea-Coulson process \be \frac{\partial}{\partial \tau} \tilde
f(m,\tau)+\frac{\partial}{\partial m}[(\gamma m+\nu \tilde
N(\tau)) \tilde f(m,\tau)]=\frac12\frac{\partial^2}{\partial
m^2}[(\gamma m+\nu \tilde N(\tau)) \tilde f(m,\tau)].
\label{eq:fp3}\ee Clearly the above Fokker-Planck approximation
preserve the mean and the variance evolution of the classical
Lea-Coulson setting.

Note that for the simplified model we obtain the standard
diffusion approximation of a Poisson process \be
\frac{\partial}{\partial \tau} \tilde f(m,\tau)+(\gamma \tilde
M(\tau)+\nu \tilde N(\tau))\frac{\partial}{\partial m}
f(m,\tau)=\frac12(\gamma \tilde M(\tau)+\nu \tilde
N(\tau))\frac{\partial^2}{\partial m^2} \tilde f(m,\tau).
\label{eq:fp2}\ee

We remark that higher order truncation can be considered as well.
For a detailed discussion of the properties of solutions of
Kramers-Moyal-expansions for a discrete Poisson process, truncated
at an arbitrary order we refer to \cite{RV}. Although positivity
of the solution is lost truncations at higher order are in better
agreement with the exact solution than the second order solution.
However the numerical solution of higher order truncations
presents some non trivial difficulties related to stability of the
solution. For such reason here we will not explore further this
direction.

Let us finally remark that equations (\ref{eq:fp1}),
(\ref{eq:fp3}) and (\ref{eq:fp2}) are of the general type \be
\frac{\partial}{\partial t} f(m,t)+\frac{\partial}{\partial
m}[(a(t)m+b(t))f(m,t)]=\frac{\partial^2}{\partial m^2}
[(c(t)m+d(t))f(m,t)]. \ee

In the case $c(t)=0$ they can be easily reduced to the heat
equation introducing the change of variables (see for example
\cite{Stn})
\begin{eqnarray*}
f(m,t)&=&u(t)g(y(t,m),\tau(t)),\qquad\quad\,\,
u(t)=\exp\left(-\int_0^t a(s)\,ds\right),\\
y(t,m)&=& u(t)m-\int_0^t b(s)u(s)\,ds,\quad \tau(t)=\int_0^t
d(s)u(s)^2\,ds,
\end{eqnarray*}
which gives \be \frac{\partial}{\partial \tau}g(y,\tau)=
\frac{\partial^2}{\partial y^2} g(y,\tau).\ee Thus from the
well-known solution to the heat equation we recover the explicit
solution for the probability density of mutants in time for
equations (\ref{eq:fp1}) and (\ref{eq:fp2}). We omit the details.

From the above considerations it is clear that approximations
(\ref{eq:fp1}), (\ref{eq:fp2}) and (\ref{eq:fp3}) may be
inadequate since the distribution of mutants, in general, is far
away from having a Gaussian shape. A possible alternative is to
replace in the mutation dynamic the Poisson process with a
different stochastic process in such a way that the scaled
mean-filed equations in the asymptotic limit originate
Fokker-Planck models of the type studied in \cite{CPT, PT1}. Such
Fokker-Planck equations cannot be reduced to the heat equation and
originate Gamma-like distributions in the long-time behavior. Here
we do not explore further this direction leaving it open for
future research.

\section{Numerical examples}
In this section we compare the continuous distribution of mutants
obtained using the different kinetic models (\ref{eq:km}) and
(\ref{eq:km2c}) in the generalized Fokker-Planck limit and some
standard methods to compute the approximated discrete distribution
in the Luria-Delbr\"ck  and Lea-Coulson settings (see Lemma 2 page
11 in \cite{Zheng}).

For the numerical solution of the kinetic master equations we
adopt a standard Monte Carlo simulation method as it usually done
in rarefied gas dynamics (see \cite{Ce, PR}). Here the main
difference is the necessity to generate Poisson samples at each
time step. This can be easily achieved by standard algorithms, see
for example \cite{Kn, Mar}.

The test cases considered has been proposed in \cite{Zheng}. We
start from an initial conditions where no mutants are present
$f_0(m)=\delta(0)$ and $N_0=1$. The parameters are $\mu=10^{-7}$,
$\alpha-\mu=3$ and the final computation time is $t=6.7$. To
reduce fluctuations the total number of simulation samples is
$5\times 10^5$. First we consider the Luria-Delbr\"uck case for
$\beta=2.5$ and then the Lea-Coulson case for $\beta=2.8$.

\begin{figure}[ht]
\begin{center}
\includegraphics[scale=0.6]{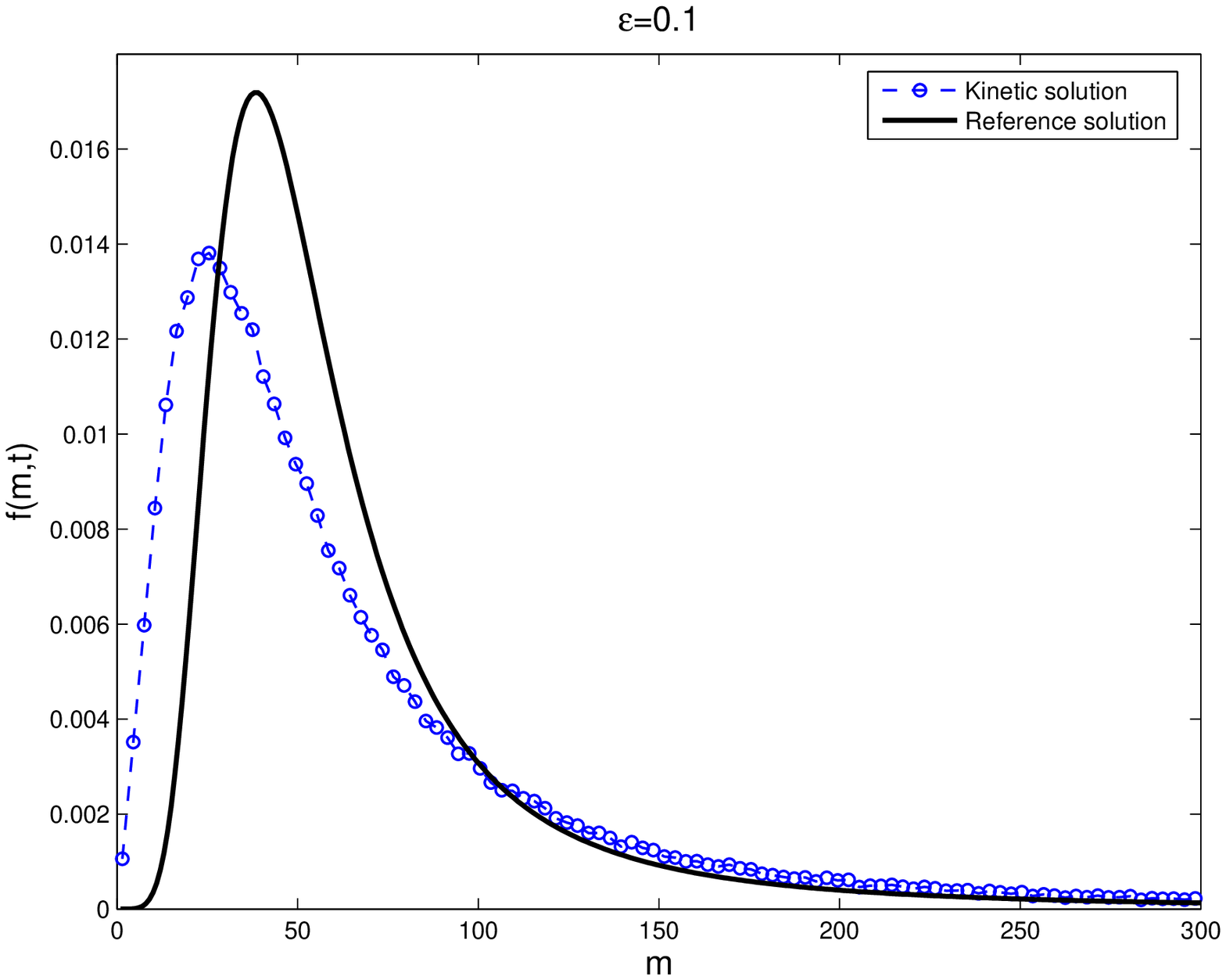}
\includegraphics[scale=0.6]{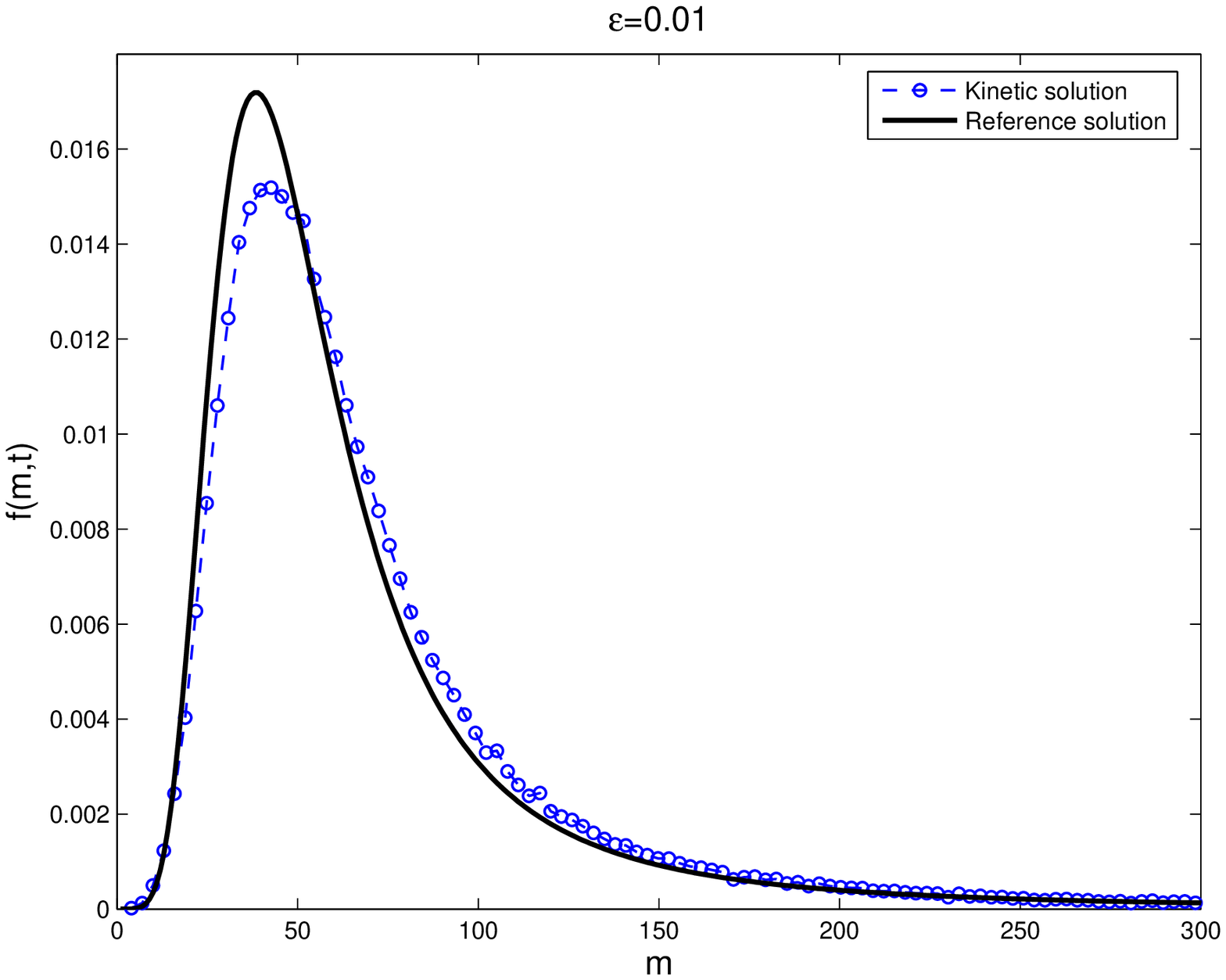}
\caption{Luria-Delbr\"uck setting. Distribution of mutant cells at
$t=6.7$ with $\beta=2.5$ for $\epsilon=0.1$ (top) and
$\epsilon=0.01$ (bottom) in the kinetic model (\ref{eq:km}). The
reference solution is computed using Lemma 2 in \cite{Zheng}.}\label{fg:k1}
\end{center}

\end{figure}

In Figure \ref{fg:k1}, we report the solution obtained simulating
the kinetic model (\ref{eq:km}) for different values of the
scaling parameter $\varepsilon$. The results show the convergence of
the mutant distribution prescribed by model towards the classical
Luria-Delbr\"uck solution obtained using Lemma 2 in \cite{Zheng}.

\begin{figure}[ht]
\begin{center}
\includegraphics[scale=0.6]{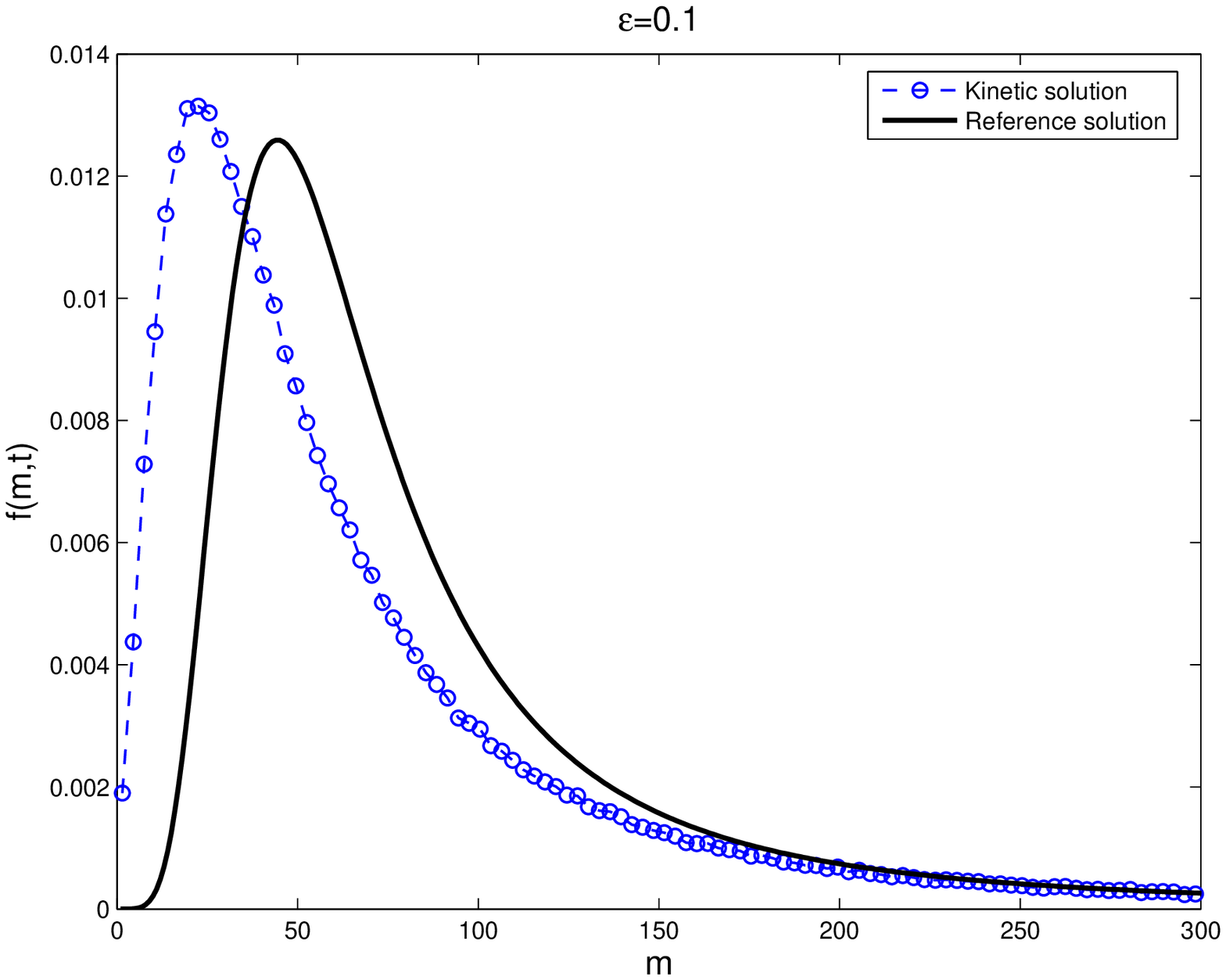}
\includegraphics[scale=0.6]{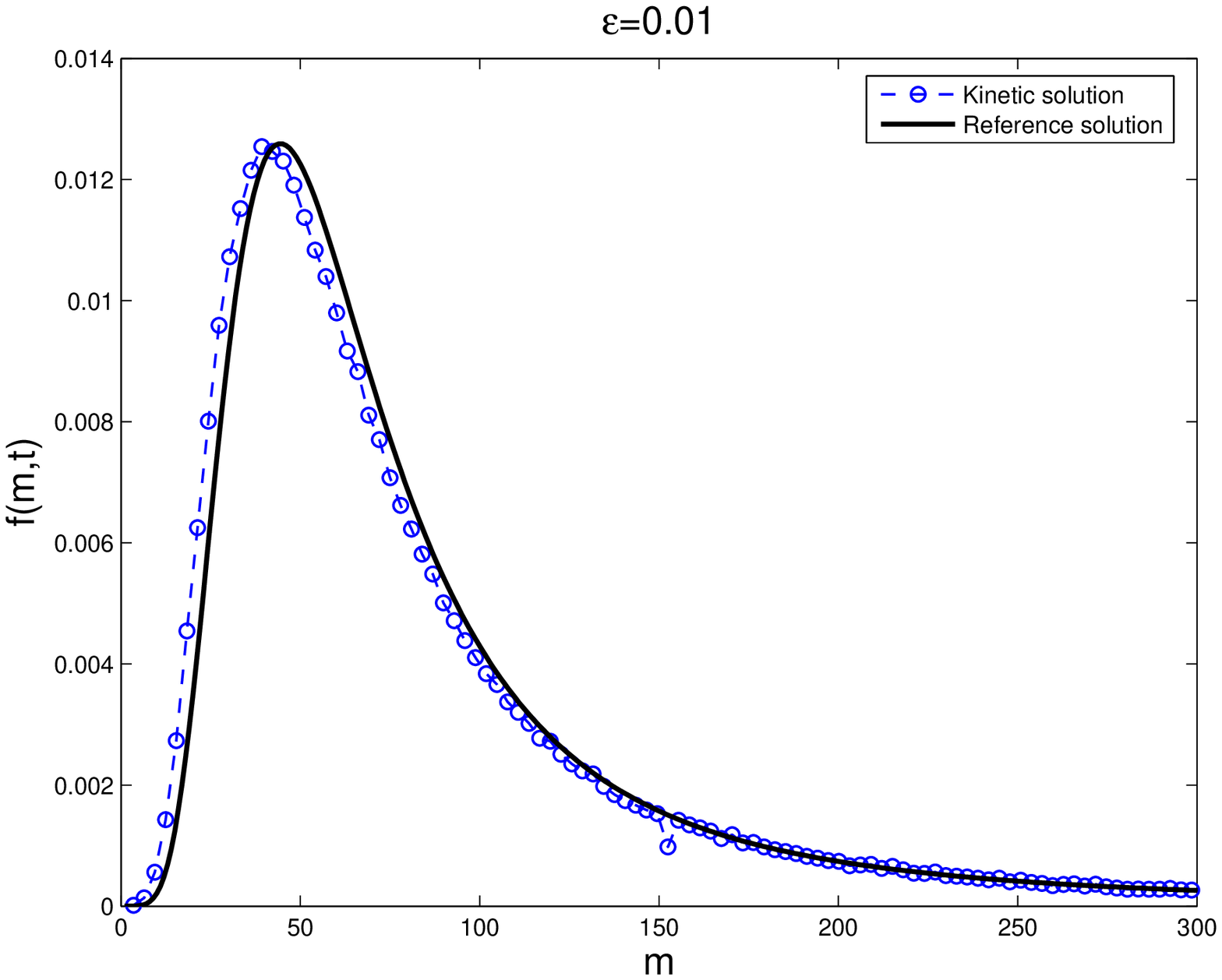}
\caption{Lea-Coulson setting. Distribution of mutant cells at
$t=6.7$ with $\beta=2.8$ for $\epsilon=0.1$ (top) and
$\epsilon=0.01$ (bottom) in the kinetic model (\ref{eq:km2c}). The
reference solution is computed using Lemma 2 in \cite{Zheng} and
numerical quadrature.}\label{fg:k2}
\end{center}

\end{figure}

Similarly we report in Figure \ref{fg:k2} the solution of the
kinetic model (\ref{eq:km2c}) for different values of the scaling
parameter $\varepsilon$. As expected convergence of the mutant
distribution towards the Lea-Coulson solution, computed using Lemma 2
together with numerical quadrature as in
\cite{Zheng}, is observed.

\section{Conclusions}
We have derived kinetic and mean field equations corresponding to two most
famous mathematical formulations of the Luria-Delbr\"uck
experiment. First we discussed the original Luria-Delbr\"uck
formulation and have shown that under a suitable mean field scaling the
master equation yields a generalized Fokker-Planck equation
having the Luria-Delbr\"uck distribution as solution. Next
we extended our approach to the Lea-Coulson case and derived the
corresponding kinetic model and asymptotic generalized
Fokker-Planck limit. Following the theoretical analysis we
presented results of numerical experiments conducted using a Monte
Carlo method. We have shown a computational evidence of
convergence of our models towards the classical formulations.

Beside the mathematical aspects of the problem, the interest in the continuous
modeling of the mutation process presented here is twofold.
On the one hand, the models
provide a new environment for the development of numerical methods
to compute mutant distribution, and, on the other hand, they
represent a first step towards the development of more realistic
mean field models where the distribution function depends also on
the physical location of the mutant cells.

\section*{Acknowledgments} E. Kashdan thanks to the Italian Institute
of High Mathematics (INdAM) for the support during his stay at
the University of Ferrara.

\end{document}